\documentclass[12pt]{article}
\begin{document}
\title{Recurrence Metrics and the Physics of Closed Time-like Curves}
\author{M.S.Modgil\footnote{msingh@iitk.ac.in}
 \and Deshdeep Sahdev\footnote{ds@iitk.ac.in}}
\date{18 May 2001}
\maketitle
\noindent
 Department of Physics, Indian Institute of
Technology,Kanpur-208016, India

 \abstract

We investigate vacuum solutions of Einstein's equations for a
universe with an $S^1$ topology of time. Such a universe behaves
like a time-machine and has geodesics which coincide with closed
time-like curves (CTCs). A system evolving along a CTC
experiences the Loschmidt velocity reversion and undergoes a recurrence
commensurate with the universal period. We indicate
why this universe is free of the causality paradoxes, usually associated
with CTCs.

\section{Introduction}

The steady proliferation of solutions containing CTCs suggests
that closure in time is more generic than esoteric in the General
Theory of Relativity (GTR) . Indeed, following Godel's \cite{God}
discovery of a universe containing CTCs, these were found to
exist in several other space-times as well, one of which in fact
predated Godel's: In 1974, Tipler \cite{Tip} showed that van
Stockum's \cite{Sto} 1936 solution of Einstein's equations, for a
rapidly rotating infinite cylinder, had CTCs (noted earlier, for
the interior, by Maitra \cite{Mai}) which allowed any two points
of this space-time to be causally connected. CTCs were
subsequently also found in the Taub-NUT \cite{Mis69},
 Kerr \cite{Car} and Tomimatsu-Sato \cite{Gib}
solutions, in worm-hole space-times \cite{Mor} and in the
neighbourhood of oppositely moving cosmic strings \cite{Gott}.
CTCs have also been discovered by Li \cite{Li} and Everett
\cite{Ever96}.

In view of this, it is difficult to maintain that the
corresponding space-times are pathological, much as any attempt
to make physical sense of them brings up a host of fundamental
(and vexed) questions of interpretation. The most important of
these are the issues of causality violation and non-unitary
evolution, to which several authors have reacted by arguing
vigourourly that in {\it no} space-time can CTCs be actually {\it
generated}: they can be at best be {\it already} present
\cite{Hawk}.

The consistency of quantum field theories defined on such
space-times has also generated some debate. Thus, while the Cauchy
problem for the scalar wave equation is well-defined on wormhole
space-times, \cite{Fre} it was pointed out that vacuum
polarization effects produce a divergence, near the Cauchy
horizon, in the renormalized stress-energy tensor of the scalar
\cite{Boul} field, but the divergence then turned out to have
nothing to do with CTCs \cite{Kras}.

Finally, quantum mechanical versions of time machines seem to fit
in well with the many-universe interpretation of quantum
mechanics \cite{Deu,Ever57} while path integrals generalized to
include those on the CTCs of a time-machine show that, for unitary
evolution, the time machine must be causally isolated from the
rest of the Universe \cite{Pol}.

Leaving aside for the moment, the bigger question of what happens
if a region containing CTCs, is {\it embedded} in a Universe with
an otherwise $R^1$-topology of time, let us turn to what exactly
happens to a system which circumnavigates a CTC (in the interior
of a time-machine, for instance)? Since none of the CTCs
discovered so far coincides with a geodesic of the underlying
space-time, motion along these occurs only under the influence of
external (i.e. non-gravitational) forces. For the Godel
space-time, for example, it requires very high accelerations and
large time-periods to return to one's past \cite{Pfa}.

The need for external forces can, however, be obviated if we are
willing to turn the entire Universe into a time-machine, i.e. to
consider the time-dimension itself to be closed. CTCs then
coincide with geodesics and the discussion can be carried out
entirely in terms of gravitational free-fall. Furthermore, since
nothing but the time-machine now exists, the question of
isolation is inapplicable and we can expect unitarity to hold
unconditionally.

Another major gain of this approach is simplicity. Recall that
for the quantum mechanical particle inside a box, the solution of
the governing Schrodinger equation is immediate. All one has to
understand the effect of the boundary conditions. Likewise, the
simplest Universe with $S^1$-time is the $S^1$ analog of
Minkowski space. But the solution contains some non-trivial
physics (due to constraints imposed by closure in time) and much
of this applies to CTCs in general.

We note, parenthetically, that while the topology of the time
dimension, for the Universe we live in, is {\it locally} known to
be $R^1$, {\it globally}, it is actually more logical to assume it
to be that of a circle whose radius must be constrained by
observation. This would be completely analogous to our pinning
down the spatial curvature of our Universe by measuring its
density and comparing it to the critical density required for
closure. It is, thus, surprising that the closest anyone has come
to doing this is Segal, who developed a model of the Universe in
which future and past infinity were identified, and in which the
standard cosmological observations found alternative explanations
\cite{Seg}.

The rest of the paper is organized as follows. Since the boundary
conditions contain much of the physics, Section II is entirely
devoted to a discussion of the time-closure constraint. In
Section III, we discuss two different vacuum solutions, a
time-independent pedagogical one in III(a), and a class of
time-dependent ones in III(b). In Section IV, all the insights
gained are put together and discussed in the context of CTCs in
general.

\section{The Time-Closure Constraint}

A Lorentzian manifold with an $S^1$ topology of time is periodic in the
sense that every event has an infinite number of copies in both past and
future. In such a universe, {\it all} particle paths --- geodesic as well
as non-geodesic --- close in an identical period and the initial Cauchy
data on a space-like hypersurface recurs an infinite number of times.

To understand what this means, consider the manifolds \cite{Zoll},
$S^n$ and $RP^n$ (i.e. $S^n$ with antipodal points identified),
{\it all} of whose geodesics close in the same proper distance
--- $2 \pi R$ and $\pi R$ respectively, where $R$ is the radius of the sphere.
Consider further an arbitrary distribution of (non-interacting) photons in
these spaces, and take the photons to be moving
along geodesics. Each photon will clearly circumnavigate the great
circle/semi-great circle it happens to be on, in the same time period,
$T (= 2 \pi R/c$ or $\pi R/c)$. As a result, the spatial
distribution of photons at time $T$, will be identical to the one we started
with. Note that for this to work, all particles
in the distribution have to be moving with the same speed. Thus if, in
addition to photons, the distribution contains material particles, recurrence
by circumnavigation will {\it not} occur.
It follows that $S^3 \times R^1$ ($S^3$-space, $R^1$-time)
and $P^3 \times R^1$ are 4-dimensional
Lorenzian manifolds, all of whose null geodesics are periodic.
(Such space-times have been studied in the literature
\cite{Gui}, and are said to have Zollfrei metrics).
\footnote{Manifolds with Zollfrei metrics automatically
regularize the infra-red divergences of
field theories defined on them \cite{Sch,Vill}.}
In the purely light-like sector, they are therefore equivalent
to $S^3 \times S^1$ ($S^3$-space, $S^1$-time)
and $P^3 \times S^1$  respectively.
\footnote{ $S^3 \times S^1$
was the basis of Segal's cosmological model \cite{Seg}.}

We turn next to a mechanism which {\it can} bring about the
equi-periodic closure of time-like geodesics in Lorentzian manifolds.
The key to this mechanism is provided by the recurrence
(i.e., the re-emergence of arbitrarily specified initial data on
a {\it distinct} space-like hypersurface \cite{Tip79}),
which this closure must inevitably produce.
We emphasize that we are talking about an {\it exact}
recurrence, occurring in a {\it finite} proper time,
and {\it not} about the arbitrarily close, asymptotic recurrence which
Poincare discussed in the context of classical ergodic systems.

Recall that what Poincare proved was this \cite{Hal74}: Given a
set $M$ of points, and  its
$\sigma$-algebra ${\cal M}$ of subsets, let $x \in C \in {\cal M}$,
and let ${\cal T}$ be a measure-preserving transformation
group of $M$ parameterized by a real and continuous $s$.
(For a statistical system, $M$ would represent
the phase space of configurations; $x$, a specific configuration, and
${\cal T}$, the group of flows parametrized by time $s$).
Then, there exists an $s$, such that ${\cal T}_s (x) \in C$, for any
$C \in {\cal M}$ and $x\in C$. In other words,
{\it given sufficient time}, the system
will return {\it arbitrarily close} to its initial state, and will do so
infinitely often.
The recurrence is clearly not exact and the theorem does not, {\it in any
way}, limit the amount of time it takes the system to return to an arbitrarily
specified neighbourhood of the initial state.

Loschmidt, on the other hand, pointed out that an {\it exact}
recurrence {\it would} occur, in a thermodynamic system,  if
each and every particle of the system underwent a {\it simultaneous}
velocity reversion, i.e, the momentum, ${\vec p}$, of {\it every}
particle changed to ${-\vec p}$ on a single time slice; put
another way, at a given instant of time, every particle (photons
inclusive) encountered an
invisible, infinitely massive wall, at normal incidence, and suffered
a reflection. Indeed, once such a velocity reversion occurs, every particle
retraces its path in configuration space, and re-enacts each collision
it had earlier suffered --- in reverse chronological order:
A film of the microscopic evolution of the system
simply starts running backwards from the moment of reversion,
the time-reversed evolution being perfectly permissible
because Newtonian dynamics is invariant under time-reversal.

We note that while a single reversion is enough to make the system
retrace its path in configuration space, {\it two} reversions
are required to make the phase space trajectory as a whole,
periodic. At each reversion, the point representing
the system's configuration in phase space, behaves as follows:
Its projection, on the subspace of momenta, jumps discontinuously
while in the (configuration) subspace of spatial positions, its projection
continuously backtracks.
Furthermore, if two such reversions occur with a given periodicity, the
system executes a cycle, limited at each end by the imaginary walls
producing the reversions. An inverse ${\cal T}_s^{-1}$ \cite{Hal56} can
then be defined for every flow ${\cal T}_t$ (and
a group character can thereby be conferred on ${\cal T}$).
Indeed, letting ${\cal C}$ denote Cauchy data at time $t$,
${\cal T}_t ({\cal C}(0)) = {\cal C}(t)$. Since, ${\cal C}(0)=
{\cal C}(T)= {\cal T}_T ({\cal C}(0))$, it follows that ${\cal T}_T = Id$.
Now, ${\cal T}_T ={\cal T}_{T+ t- t} = {\cal T}_t {\cal T}_{T - t} , \Rightarrow
{\cal T}_t^{-1} = {\cal T}_{T - t}$.

For a {\it classical} thermodynamic system, there is, however, no
mechanism for effecting such reversion in velocity and Loschmidt's
observation has thus remained a mere curiosity.
In GTR, on the other hand, such a mechanism {\it does} exist, as we show
in detail below. The reason is that here,
particle systems are tied to geodesics which, in turn, are determined
by the underlying metric. If this happens to be a solution to the
Einstein field equations on a manifold with $S^1$ time, recurrence
is enforced by the metric itself, i.e. the corresponding geodesics
automatically incorporate the Loschmidt velocity reversion.

Explained differently,
$S^1$-time introduces an effective velocity-dependent potential
which reverses the trajectories of all particles on a single time slice.
The situation is somewhat analogous to that of a particle thrown up
vertically in the earth's
gravitational field (with less than the escape velocity):
The particle eventually stops, reverses and reacclerates downwards.
In doing so, it revisits all the positions it had previously occupied,
at the corresponding speeds, before reaching its starting point. The
time of reversion in this case is however dependent on the intial
velocity, in contrast to what is implemented by $S^1$-time.

Two remarks are in order at this point. Firstly, the above result
does not, in any way, contradict the `no return' theorem \cite{Tip79},
proved  by Tipler, for a closed universe, beginning and ending with a
singularity, simply because the premises of that theorem do not
hold for a universe with $S^1$-time. Secondly, since a particle's position
in GTR is represented by a 4-vector, one of whose components is time,
we intuitively expect that the backtracking in space
brought about by (4-)velocity reversion, would in general be accompanied by
backtracking in time, i.e. by time-reversal. In other words, we expect
the discussion to be closely tied to the circumnavigation of a CTC, as
indeed it is.

To see the time reversal explicitly, we first note that there are 2
distinct times in the problem: The coordinate time, $t$, and the proper
time, $\tau$. The former appears in the metric and takes on all values
on the real line. It wraps around the $S^1$ of time once, each time
it increases by $T$, the period of recurrence.
The latter, on the other hand, is the time kept by a clock comoving with
a particle inside the periodic universe. It is {\it this} which displays
the reversal we are discussing.

The two times can be simply related
by setting all spatial separations in the line element to 0 and
identifying the proper distance with the proper time to get
$d\tau=\sqrt{g_{{}_{00}}}dt$. Now for a universe with $S^1$-time,
$g_{{}_{00}}$ is necessarily periodic and smooth in $t$, which
makes $\tau$ likewise so. But then $\tau$ must have 2 (or, more generally,
an even number
\footnote{
Cf Milnor \cite{Mil} for a proof based on Morse Theory of the following
result: A continuous function $f$ defined on a closed curve $\Gamma$
parametrized by $s$, has an even number of critical points where $df/ds=0$.}
of) turning points in the interval $[0, T]$. This, in turn, implies that
every time-cycle {\it must} have a region of time-reversal where
$d\tau/dt < 0$, i.e. $\tau$
decreases monotonically, and {\it all} processes reverse direction,
bringing about, among other things, a decrease in entropy. It is,
incidentally, the existence of this region that enables {\it all} physical
clocks to display the same readings at coordinate times, $t=0$ and $t=T$.

Interestingly, this time-reversal seems to be completely generic.
Its occurrence has been {\it individually} noted for each
of the space-times in which CTCs have been found to exist.
We shall return to this point in our concluding discussion.

\section{Some Vacuum Solutions}

We now examine some explicit examples of recurrence metrics satisfying the
vacuum Einstein equations. The first of these has the virtue
of being mathematically trivial, and yet capturing the physical essence of the
problem.

\subsection{The Periodic Minkowski Space-time}

Consider the following modification of the Minkowski metric:
\begin{equation}
d s^2 = c^2 d[F_T (t)]^2 - dx_1^2 - dx_2^2 - dx_3^2 ,
\end{equation}
\noindent
where the function $F_T (t)$

\begin{itemize}
\item is periodic in $[-T/2,T/2]$, and
\item satisfies $\lim_{T\rightarrow\infty} F_T (t)=t$,
\end{itemize}
a concrete example being
\begin{equation}
F_T (t) = \frac{T}{2\pi} \sin \frac{2\pi t}{T}.
\end{equation}

The Minkowski space-time (or, more precisely, the union of this and the point
at infinity) can be recovered as a limiting case of this
(periodic) universe: We simply let the recurrence period $T\rightarrow\infty$,
and note that $F_T(t)$ must then coincide identically with $t$.

The only non-zero Christoffel symbol corresponding to the above metric is
\begin{equation}
\Gamma^0_{00} = {\ddot {F}_T (t)}/{\dot{F}_T (t)}.
\end{equation}

This does not affect, in any way, the Riemann curvature tensor:
All components of the latter vanish identically, as for the Minkowski
metric. The periodic Minkowski space-time is thus flat and constitutes
a vacuum solution of the Einstein Field Equations.

$\Gamma^0_{00}$ does however alter the geodesic equations,
which are now given by
\begin{eqnarray}
\nonumber t''(s) - t'(s) \frac{\ddot{F}_T (t)}{\dot{F}_T (t)} = 0,\\
 x''(s) = 0, i=1,2,3
\end{eqnarray}
\noindent
where primes and dots denote differentiations with respect to the geodesic
parameter, $s$, and the time, $t$, respectively.

These equations readily integrate to
\begin{eqnarray}
\nonumber  t &=& F^{-1} [c_1^0(s-c_2^0)],\\
x_i &=& c_1^i s + c_2^i , i=1,2,3,
\end{eqnarray}
\noindent
where $c_1^\mu$ and $c_2^\mu (\mu= 0,1,2,3)$ are integration constants.
These can be further combined to express the $x_i$ as functions of $t$:
\begin{eqnarray}
\nonumber x_i(t) &=& a_i F_T (t) + b_i,\\
a_i &=& {c_1^i}/{c_1^0},\\
b_i &=& c_1^i c_2^0 + c_2^i.
\end{eqnarray}
\noindent
For a stationary particle, $a_i = 0$, for all $i$. For a
photon propagating in the $x_1$-direction, say, $a_1 = c$, and
$a_2 = a_3 = 0$. For these values of the $a_i$, the line element
is trivially seen to vanish.

We now note that a velocity reversion, marked by $d x_i/d t=0$,
occurs whenever $\dot{F}_T (t)=0$.
The periodicity of $F_T (t)$ further guarantees that
this condition is fulfilled at a minimum of 2 points in every cycle,
the corresponding times being
{\it independent} of the values of $a_i$ and $b_i$. In other words,
each reversion occurs at the {\it same} value of $t$
on {\it every} geodesic. These reversions are thus clearly of
the Loschmidt kind.

What is more,
\begin{itemize}
\item The proper time, $\tau = F_T (t)$, reverses at these points as well,
making moments of velocity reversion coincide with those of T-reversal.
Thus $\tau$, in contrast to the coordinate time $t$, does {\it not} increase
monotonically within the interval $[-T/2,T/2]$. Rather, in the simplest case
of only 2 points of reversion, $t_1$ and $t_2$ say, it increases from  $t_1$
to $t_2$, and decreases over the rest of the cycle.

\item The line element becomes purely space-like at the points of
velocity reversion. This is reflected in the behaviour of the light cone,
whose semi-angle, $\alpha=\arctan \dot{F}_T(t)$, vanishes at these points,
making the entire space-time {\it momentarily} space-like.

\item Each point of the periodic Minkowski space-time has a horizon of
radius, $r= c F_T^{max}$ (where $F_T^{max}$ is the maximum value of $F_T$):
No particle --- massive or massless --- can go to, and no information
can come from, {\it any} point beyond the horizon.
Our metric thus acts like/produces a gravitational potential well,
confining all particles and photons to within this horizon.
This situation is analogous to that occuring within the Schwarzchild
radius of a black-hole: Photons moving radially outwards are
dragged back. Our space-time is of course singularity-free
but nonetheless the escape velocity is infinite.
\end{itemize}

Thus if $F_T$ is of the form given in Eq.(2), with $T$ set to $2\pi$
for simplicity, $\tau = F_{T=2\pi}(t) = \sin t$, $\alpha =
\arctan cos t$, and the horizon radius $r = c$. The time cycle
extends from -$\pi$ to +$\pi$, and velocity reversions
occur at $\pm \pi/2$. The proper time $\tau$ increases, as a function
of $t$, from -$\pi/2$ to +$\pi/2$, and decreases in the remaining
part of the cycle. The light cone contracts from a semi-opening angle
of -$2 \times \arctan 1$ at $t = -\pi \equiv +\pi$ to 0 at $t = -\pi/2$,
reopens (in the opposite direction, so to say) out
to an angle of +$2 \times \arctan 1$ at $t = 0$, closes again to 0 at
$t = \pi/2$ and finally returns to its initial value at $t = \pi$.

All this is for non-interacting particles. What happens if
we turn interactions on? The question is difficult to answer in
general. However, if we take the interactions to be point-like,
a particle would simply switch from one geodesic to another,
each time it undergoes an interaction. Subsequent to a moment of
velocity reversion, all particles would retrace their paths and collisions.
and undergo recurrence exactly as non-interacting particles would.

\subsection{A Simple Generalization}

It is not difficult to construct more general universes with an
$S^1$-topology of time. For example, a fairly rich class of recurrence
metrics is obtained if we make the spatial components cyclic as well
(with the Universal period, T). The corresponding line element is then of
the form,
\begin{equation}
d s^2 = c^2 d F_T(t)^2 - d F_T^1 (t,x_1)^2 - d F_T^2 (t,x_2)^2 -
d F_T^3 (t,x_2)^2
\end{equation}
where the functions, $F_T^i$,
\begin{itemize}
\item Have dimensions of length,
\item Are periodic in $[-T/2,T/2]$,
\item Satisfy $lim_{T\rightarrow\infty} F_T^i (x_i) = x_i,   i=1,2,3$.
\end{itemize}

It can be readily verified that the above metric yields a vacuum
solution of the Einstein field equations. Moreover, the $t$-component of
the geodesic equations is same as for the periodic
Minkowski metric. The spatial components are, however,
much more complicated and are, in fact, not integrable for arbitrary $F_T^i$.

Since these technically more involved solutions are well outside the
primary focus of this paper, we shall not pursue them any further at
this point.

\section{Summary and Discussion}

To summarize, we have examined a new class of vacuum solutions to
Einstein's equations, corresponding to (recurrence) Universes with
an $S^1$-topology of time. Each point of these space-times lies on
a geodesic CTC. In contrast to say the Godel and von Stockum universes,
which also have CTCs passing through all points, recurrence Universes
are free of rotations and simple enough to permit a detailed
analysis of how closure in time actually works.

Our analysis has revealed that in recurrence space-times, particle
systems return
to whichever state they started from at the end of a universal cycle
of period $T$, as a result of an even
number (per cycle) of reversions in the velocities of all constituent
particles. The velocities become zero on a hypersurface characterized
by a single value of coordinate time, and reverse
without discontinuity. Once they do so, the system begins to revisit
the states it had earlier occupied --- in reverse chronological order.
This means that if the entropy was earlier increasing, as in the
mixing of two confined gases initially separated by a diaphragm,
it now begins to decrease, i.e. the gases begin to separate out
(and {\it vice versa}). In other words,
each reversion results in a reversal of the arrow of time, defined
through entropy change.

This adds an unexpected twist to our understanding of how
time-reversible equations at a microscopic level produce time-irreversibility
on macroscopic scales. Indeed they seem to suggest that the breaking of
T-invariance is, in some sense, spontaneous: There actually exist two
distinct phases --- one of entropy increase and one of entropy decrease
but only one of these is realized in any given situation. In Universes
with $R^1$-time, there is no way of going from one to the other, but
on a CTC there is. In fact, one goes continuously from one to the other,
without an energy barrier, as in a second-order phase transition.

We have also found that each reversion is marked by a reversal
in the direction of proper time with respect to the time coordinate,
parametrizing the temporal $S^1$. We have, in turn, traced this to
the fact that proper time is periodic (with period $T$) and smooth,
on all CTCs.  Indeed, to be so, as a funtion
of the linearly increasing coordinate time, it must have an even number
of turning-points on every CTC, each of which must, therefore, contain
regions of negative proper time evolution.

This is true, as briefly mentioned, not merely for the CTCs of
recurrence space-times but for CTCs in general. In the Godel
universe, for example, all time- and light-like geodesics are
bounded by a horizon \cite{Nov}, which is however pierced by CTCs.
Negative time travel occurs as one crosses this horizon (with the
help, necessarily, of a large non-gravitational force). In
wormhole space-times, CTCs thread the wormhole and one goes
backwards in time while traversing the latter's throat. In the
rotating Kerr, Tomimatso and von Stockum solutions, the $g_{\phi
\phi}$-component of the metric changes sign for certain values of
the coordinates, making $\phi$ time-like. If we now move in the
$-\phi$-direction, we begin moving backwards in time. The
situation in a Gott universe (where we have a deficit wedge angle
between the two halves into which the plane passing through two
parallel moving cosmic strings divides 3-space) is more
complicated. The CTCs of this space-time pass through the deficit
wedge region and it is here that negative time evolution occurs.

How does the simultaneous occurrence of negative proper time
evolution and entropy decrease tie in with our psychological sense of time?
We know that processes evolve locally according to proper time, that they
must always be timed one against another, and that psychological
time results from watching irreversible processes in the backdrop
of simple, reversible oscillatory ones.
(These are incidentally not affected by the universal velocity reversion,
or rather, they keep undergoing reversions on a much shorter but
commensurate time-scale).  Thus if each conventionally irreversible
process reverses, the sensation will be one of moving backwards in time.

This assumes however that the observer's memory, as a system, is not
being continuously
reset in the process. The universal evolution of {\it all} systems towards
states they earlier occupied suggests that this assumption should probably
be revised.

If we do this,
then at $t_2+\Delta t$, in a cycle containing reversions at $t_1$ and
$t_2$, with entropy increasing in $[t_1, t_2]$ and decreasing in $[t_2, t_1]$,
all systems, {\it including the observer's memory} will revert to the states
they
occupied at $t_2 - \Delta t$. But at $t_2 - \Delta t$, the information
stored in the observer's memory relates exclusively to what occurred
in $[t_1, t_2 - \Delta t]$. This means that in evolving forward in $t$,
from $t_2$ to $t_2 + \Delta t$, all memory of what occurred between
$t_2 - \Delta t$ and $t_2 + \Delta t$ is effaced. In other words, at
every point in the cycle, the observer is aware only of what has occurred
in his conventional past, and if proper time evolves backwards, this
past keeps shrinking.
This moreover suggests that entropy decrease may actually be unobservable
because, at no stage, is the memory of any state with entropy higher than
the present ever retained.

In view of the necessity to retrace previously occupied states, it is
tempting to speculate that the resolution of the grandfather paradox
lies in keeping careful track of the states of the observer as he moves
back in time along a CTC. For he must grow younger as he
moves back in time towards his childhood and eventually his conception.
As such, he moves {\it only} into his own past and can never access a point
in time prior to his birth.  The paradox arises from the erroneous assumption
that the observer can move back in time without himself changing in
any way.

We conclude by raising some issues, which, while being outside the
scope of the present discussion, constitute important directions for
future research. Firstly, the cosmology of a
universe with $S^1$-time is of interest, particularly with a
view to putting an observational bound on the value of $T$.
Secondly, the quantum generalizations of
arguments given in this paper, at the classical level, are essential
for a resolution, among other things, of questions relating to non-unitary
evolution on CTCs. Thirdly,
it is important, for our understanding of time-machines, to extend
everything we have discussed to other space-times with CTCs.
These and other related topics will be reported on elsewhere.

\section{Acknowledgement}
The authors would like to thank V. Ravishankar for several enlightening
discussions.

\end{document}